\documentclass[twocolumn,amsmath,amssymb]{snp}
\usepackage{graphicx}
\usepackage{dcolumn}
\usepackage{bm}
\begin{document}

\title{{\Large Skyrme-force dependence of fusion cross-sections using Wong formula in semiclassical formalism}}

\author{\large Raj Kumar}
\author{\large Raj K. Gupta}
\affiliation{Physics Department, Panjab University, Chandigarh-160014, INDIA.}

\maketitle

\section*{Introduction}
In another contribution to this conference \cite{bansal10} and Ref. \cite{kumar09}, we have shown that, for use of the
nuclear proximity potential \cite{blocki77}, and with multipole deformations up to hexadecapole ($\beta_4$), the angular
momentum $\ell$-summed Wong formula is sufficient to explain the capture cross-sections for $^{48}$Ca+$^{238}$U, $^{244}$Pu
and $^{248}$Cm reactions forming superheavy nuclei, but need (additional) modifications of barriers at sub-barrier energies
for the fusion-evaporation cross-sections in $^{58}$Ni+$^{58}$Ni, $^{64}$Ni+$^{64}$Ni, and $^{64}$Ni+$^{100}$Mo reactions
known for fusion hindrance phenomenon in coupled-channels calculations. Some barrier modification effects are shown to be
already present in $\ell$-summed Wong expression though its inbuilt $\ell$-dependence \cite{kumar09}. The non-coplanarity
degree of freedom (azimuthal angle $\Phi\ne$0) is also included \cite{bansal10}, which does not influence the fits obtained
in above noted reactions, expect for one-to-two additional units of $\ell_{max}$-value. The barrier modification effects
are also supported by the dynamical cluster-decay model (DCM) of preformed clusters by Gupta and collaborators
\cite{gupta10}, where ``barrier lowering" at sub-barrier energies arise in a natural way in its fitting of the only
parameter of the model, the neck-length parameter.

In this paper, we use the nuclear proximity potential obtained recently \cite{gupta09} for the Skyrme nucleus-nucleus
interaction in the semiclassical extended Thomas Fermi (ETF) approach using Skyrme energy density formalism (SEDF) under
frozen-density (sudden-without-exchange) approximation, and $\Phi$=0$^0$. This method allows the barrier modifications by
using different Skyrme forces, since a different Skyrme force would mean different barrier characteristics, and possibly a
different force for different reaction cross-section.

\section*{Methodolgy}
The nucleus-nucleus interaction potential in SEDF, under slab approximation of giving the nuclear proximity potential
\cite{chatto84}, is
\begin{small}
\begin{eqnarray}
V_N(R)&=&2\pi\bar{R}\int_{s_0}^{\infty}{e(s)ds}\nonumber\\
&=&2\pi\bar{R}\int \{ H(\rho)-[H_1(\rho_1)+H_2(\rho_2)]\} dz, \nonumber\\
\label{eq:1}
\end{eqnarray}
\end{small}
\vspace{-0.7cm}
\par\noindent
with $H$ as the Skyrme Hamiltonian density in ETF model with both the kinetic energy $\tau$ and spin densities $\vec{J}$
as functions of $\rho$, the nuclear density, included here up to second order, and under the frozen approximation, as
\begin{small}
\begin{equation}
\tau(\rho)=\tau_1(\rho_1)+\tau_2(\rho_2),~~\vec{J}=\vec{J}_1(\rho_1)+\vec{J}_2(\rho_2),
\label{eq:2}
\end{equation}
\end{small}
\vspace{-0.5cm}
\par\noindent
with $\rho_i=\rho_{in}+\rho_{ip}$, $\tau_i(\rho_i)=\tau_{in}(\rho_{in})+\tau_{ip}(\rho_{ip})$, and
$\vec{J}_i(\rho_i)=\vec{J}_{in}(\rho_{in})+\vec{J}_{ip}(\rho_{ip}).$ In Eq.(\ref{eq:1}),
$R=R_{1}(\alpha_1)+R_{2}(\alpha_2)+s$, where $R_{1}$ and $R_{2}$ are the temperature (T) dependent radii of two deformed
and oriented nuclei, separated by $s$ with a minimum $s_0$-value. For $\rho_i$, we use the T-dependent Fermi density
distribution
\begin{small}
\begin{equation}
\rho_i(Z_i)=\rho_{0i}(T)\left[1+\exp{\frac{z_i-R_{i}(T)}{a_{i}(T)}}\right]^{-1},
\label{eq:3}
\end{equation}
\end{small}
\par\noindent
with $-\infty \le z\le \infty$, $z_2=R-z_1$ and $\rho_{0i}(T)=\frac{3A_i}{4 \pi R_{i}^3(T)} \left[1+\frac{\pi^2
a^2_i(T)}{R_{i}^2(T)}\right]^{-1}$, and nucleon densities $\rho_{iq}$ further defined as
$$\rho_{in}=({N_i}/{A_i})\rho_i,~~\rho_{ip}=({Z_i}/{A_i})\rho_i.$$
\begin{figure*}
\vspace{-1.2cm}
\includegraphics[scale=0.99]{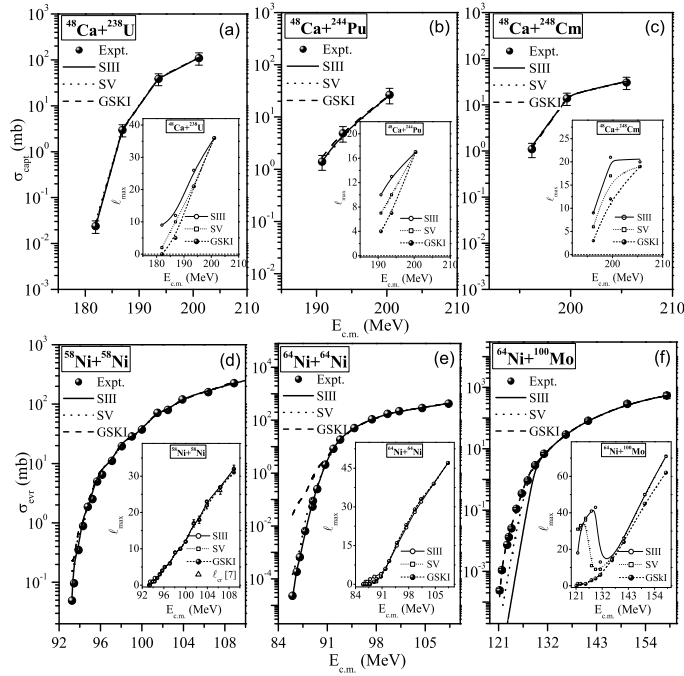}
\vspace{-1.5cm}
\caption{\label{fig1} Comparison of the calculated cross-sections using Skyrme forces SIII, SV and GSkI with experimental
data for Ca- and Ni-based reactions. Insets show the variation of deduced $\ell_{max}(E_{c.m.})$.}
\end{figure*}
Adding Coulomb and centrifugal terms to $V_N(R)$, we get the total interaction potential for deformed and oriented nuclei,
as
\begin{small}
\begin{eqnarray}
V_{\ell}(R)&=&V_N(R,A_{i},\beta_{\lambda i},T,\theta_{i})+V_{C}(R,Z_{i},\beta_{\lambda i},T,\theta_{i}) \nonumber \\
&&+V_{\ell}(R,Z_{i},\beta_{\lambda i},T,\theta_{i}),
\label {eq:4}
\end{eqnarray}
\end{small}
\vspace{-0.3cm}
\par\noindent
with non-sticking moment-of-inertia (=$\mu R^2$, with $\mu$ as the reduced mass) for $V_{\ell}$.

The $\ell$-summed Wong's formula is \cite{kumar09},
\begin{small}
\begin{equation}
\sigma (E_{c.m.},\theta_i)=\frac{\pi}{k^2}\sum_{\ell=0}^{\ell_{max}}(2\ell+1)P_{\ell}(E_{c.m.},\theta_i),
\label{eq:5}
\end{equation}
\end{small}
\par\noindent
where $k=\sqrt{\frac{2\mu E_{c.m.}}{\hbar^2}}$, and penetrability $P_{\ell}$ determined in Hill-Wheeler approximation.
Integrating over the angles $\theta_i$, we get $\sigma(E_{c.m.})$.

\section*{Calculations and Results}
Fig. 1 shows our calculations for the Ca- and Ni-based reactions, using the Skyrme forces SIII, SV and GSKI, compared with
experimental data. Apparently, the capture cross-section in Fig. 1(a) to (c) are nicely reproduced for all the forces
employed, with a small change in $\ell_{max}$-values (see insets), whereas the evaporation-residue cross-sections in Figs.
1(d) to (f) require  different forces. The $\sigma_{evr}$ for $^{58,64}$Ni+$^{58,64}$Ni are best fitted for SIII force,
and for $^{64}$Ni+$^{100}$Mo by GSkI force, but {\it not} the vice-versa, i.e., then the barrier modification is to be
added empirically. Note that the deduced $\ell_{max}$ in $^{58}$Ni+$^{58}$Ni reaction for fitted force compare nicely with
experimental values \cite{beckerman81} at higher $E_{c.m.}$'s.

Thus, for $\ell$-summed Wong formula, the Skyrme force dependence in SEDF using ETF method is evident for evaporation
residue and not for capture cross-sections.

%



\begin{thebibliography}{50}
\bibitem{bansal10}
M. Bansal and R. K. Gupta, Contribution to this Symposium.
\bibitem{kumar09}
R. Kumar, M. Bansal, S. K. Arun and R. K. Gupta, Phys. Rev. C {\bf 80} (2009) 034618.
\bibitem{blocki77}
J. Blocki, {\it et al.}, Ann. Phys. (N.Y.) {\bf 105} (1977) 427.
\bibitem{gupta10}
R. K. Gupta, Lecture Notes in Physics 818, ``Clusters in Nuclei", Vol. I, ed C. Beck, (Springer Verlag) (2010) p. 223.
\bibitem{gupta09}
R. K. Gupta, {\it et al.}, J. Phys. G: Nucl. Part. Phys. {\bf 36} (2009) 075104.
\bibitem{chatto84}
P. Chattopadhyay and R.K. Gupta, Phys. Rev. C {\bf30} (1984) 1191.
\bibitem{beckerman81}
M. Beckerman {\it et al.}, Phys. Rev. C {\bf 23}, 1581 (1981).
\end{thebibliography}
\end{document}